# Methylammonium Lead Trihalide Perovskite Solar Cell Semiconductors Are Not Organometallic

Pradeep R. Varadwaj[*]

Department of Chemical System Engineering, School of Engineering, The University of Tokyo
7-3-1, Hongo, Bunkyo-ku, Japan 113-8656
[b]CREST-JST, 7 Gobancho, Chiyoda-ku, Tokyo, Japan 102-0076

Methylammonium lead trihalide perovskite solar cells ($CH_3NH_3PbY_3$, where $Y = I_{(3-x)}Br_{x=1-3}$, $I_{(3-x)}Cl_{x=1-3}$, $Br_{(3-x)}Cl_{x=1-3}$, and $IBrCl$) are photonic semiconductors. Researches on various fundamental and technological aspects of these materials are extensively on-going to make them stable environmentally and for commercialization. Research studies addressing these materials as organometallic are massively and repeatedly appearing in very reputable, highly cited and high impact peer-reviewed research publications, including, for example, Energy and Environmental Science, Nature Chemistry, Nature Communication, Advanced Materials, Science, ACS Nano, and many other chemistry and materials based journals of the Wiley, Elsevier, Springer and Macmillan Publishers, and the Royal and American Society Sciences. Herein, we candidly addresses the question: whether should scientists in the perovskite and nanomaterials science communities refer $CH_3NH_3PbY_3$ and their mixed halogen derivatives as organometallic?

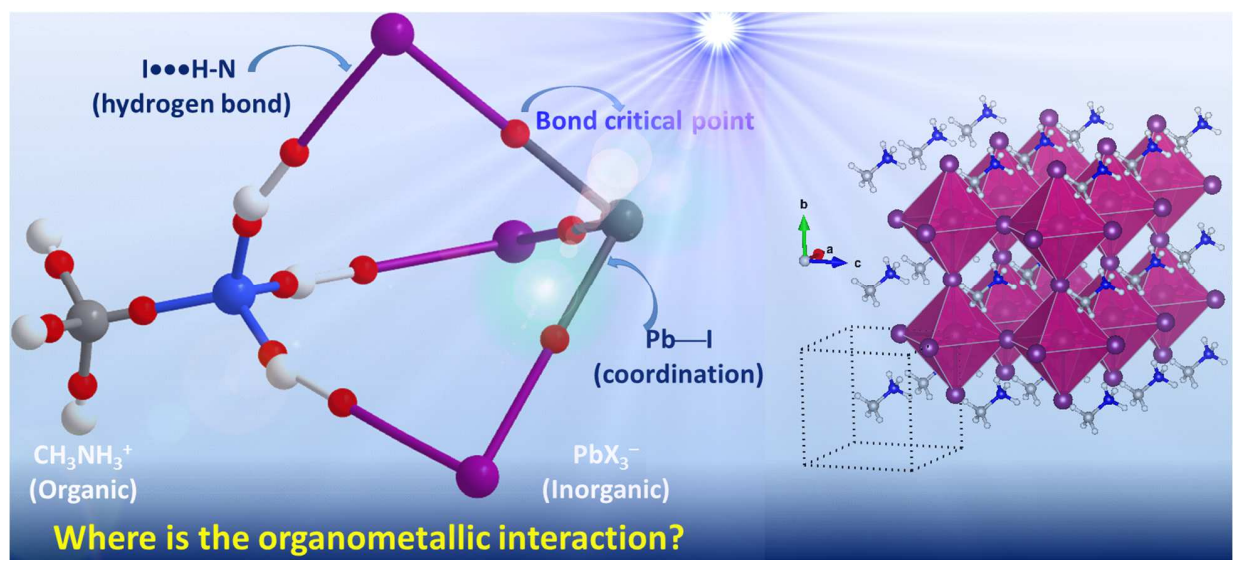

[*] Corresponding Author's E-mail Addresses: pradeep@t.okayama-u.ac.jp; pradeep@tcl.t.u-tokyo.ac.jp

Methylammonium lead trihalide perovskites (CH$_3$NH$_3$PbY$_3$, where Y = I$_{(3-x)}$Br$_{x=1-3}$, I$_{(3-x)}$Cl$_{x=1-3}$, Br$_{(3-x)}$Cl$_{x=1-3}$, and IBrCl) are splendiferous soft photonic nanomaterials.[1] These are innovative super-ion inspired man-made semiconductors, which are supremely efficient to convert light energy into electricity.[2]

The unified experimental synthetic technology developed for fabricating the CH$_3$NH$_3$PbY$_3$ solar cell materials is extremely simple. The technology can be created in ordinary research labs inexpensively as the raw materials are solution processable, cheap and abundant.[3-4]

Because of their phenomenal photon power conversion efficiency (PCE), the CH$_3$NH$_3$PbY$_3$ solar cells stand out among all other solar cells discovered till date, thereby generating a historical breakthrough in the photovoltaic solar energy technology.[5-6] For instance, the PCE for methylammonium lead triiodide (CH$_3$NH$_3$PbI$_3$, also known as MAPbI$_3$) has grown from an initially reported value of 3.8% to a National Renewable Energy Laboratory certified value of 22.1% within a few years. Such a growth of the PCE has actually taken decades for other photovoltaic solar cells to accomplish.[5-6] Nevertheless, the extraordinary and highly innovative design principles and synthetic routes discovered during the study of the CH$_3$NH$_3$PbY$_3$ and their halogen derivatives are suggested to assist the invention of the next generation trihalide based perovskite solar cell materials.[6-7] The mechanisms such as the 'ion migration',[8a] 'defect tolerance and migration',[8b-c] and several others[9] are also introduced to this research field for the fundamental understanding of the stability, device based performance, and environmental degradation mechanisms associated with these solar cells.[10,11]

The scientific community has witnessed the spectacular research advancements of the area of trihalide-based photovoltaic perovskite solar cells within a very short time. Not only this, but it also witnessed the frequent use of a scientific terminology '*organometallic or organometal*' to address these solar cell systems. It was actually in 2009 Kojima *et al.* have introduced this terminology to the scientific community by incorporating the phrase '*organometal halide perovskites …*' in the title of their paper.[5b] This interesting paper appeared in the Journal of American Chemical Society is highly cited. The research summarized in it is impressively one of the origins that has motivated many researchers to conduct analogues studies since it included a PCE of 3.8% for what is recognized today as the highest performance solution processable perovskite solar cell CH$_3$NH$_3$PbI$_3$, thus leading to many research articles published in later years. The use of the suggested terminology continues to appear vigorously, which is evident of many titles, topics and texts of the massively appeared peer-reviewed research publications of the high impact journals, such as Energy and Environmental Science, [11-12] Nature Chemistry, Photonics and Communications, [13-14] Advanced Materials, [15-16] Science, [34,17] Journal of the American Chemical Society,[18] and many other chemistry and materials based Journals of the Wiley, Elsevier, Springer, Macmillan Publishers, and the Royal and American Society Sciences.[3-4,7-8,11-18] Neither the reviewers, nor the authors put emphasis on majority of these published peer-reviewed research papers to identify and analyze which part of the $CH_3NH_3PbY_3$ system is organometallic, and which parts of it are an inorganic and an organic-inorganic.

To validate the statement above, let us now look at the histograms depicted in Figs. 1a) and b) that are generated using the Web of Science database.[19] These suggest that there is undoubtedly an upsurge of scientific interest in the halide-based perovskite solar cell research since 2009. Fig. 1a shows there are roughly 470 peer-reviewed articles published conceiving the phrase "halide perovskite" in the TITLE. The above number is redirected to 3212 when the same phrase was searched as TOPIC (cf. Fig. 1b)). As the histograms in Figs. 1c) and d) illustrate, such numbers became 110 and 1667 when the phrase 'organometal halide' was searched as TITLE and TOPIC, respectively. Also, these graphs forecast the fact that the frequency of the largest number of peer-reviewed articles published is in the year 2016, which is approximately 770 (cf. Fig. 7d)), with 68722 total citations (within the April 14, 2009 – March 15, 2017 period). Not only this, but the statistical result suggests that the specific phrase 'organometal halide' was numerously and continuously invoked in the main body of these peer-reviewed articles to address the $CH_3NH_3PbX_3$ perovskites and their mixed halogen derivatives.

In the other instance, some authors refer the $CH_3NH_3PbY_3$ perovskites to as *organometallic*. Providing evidences, we include a few peer-reviewed published articles that are having the titles such as 'Electroluminescence from *organometallic* lead halide perovskite-conjugated polymer diodes', [20] 'Two-photon absorption in *organometallic* bromide perovskites', [21] 'Deciphering halogen competition in *organometallic* halide perovskite growth', [22] 'Third-order optical nonlinearities in *organometallic* methylammonium lead iodide perovskite thin films', [23] 'Organometallic perovskite metasurfaces',[15] '*Organometallic* halide perovskites: Sharp optical absorption edge and its relation to photovoltaic performance',[24] 'Turning a disadvantage into an advantage: synthesizing high-quality *organometallic* halide perovskite nanosheet arrays for humidity sensors,[25] and 'Highly reproducible *organometallic* halide perovskite microdevices based on top-down lithography',[26] etc.

For curiosity, we have also performed another survey using the same database,[19] but this time with the phrase "*organometallic halide perovskite*". An interest towards this end was to see how many times this phrase is invoked in the peer-reviewed articles published till date to address the $CH_3NH_3PbX_3$ perovskites and their mixed halogen derivatives. Interestingly, the resulting number of articles resulted with this phrase were ca. 11 and 73 when we constricted the search options to TITLE and TOPIC, respectively. And, when this phrase was changed to '*organometallic perovskite*', and researched, the number of articles increased to become 24 and 97, respectively. Conspicuously, going through the titles of all these 35 (11 + 24) peer-reviewed articles the immediate impression a general chemist gains lies in the fact that the nature of the chemical bonding environment in the $CH_3NH_3PbY_3$ hybrid perovskite solar cells is purely *organometallic*. This impression is just not limited to only chemists. It is quite apparent that this concept confuses greatly to the graduates, undergraduates, and non-chemists.

The concept that *the $CH_3NH_3PbX_3$ perovskite solar cells are organometallic* is already propagated enough through transfer of knowledge based technologies, thereby crossing the boundary of peer-reviewed research articles. This means the concept is just not condensed within peer-reviewed journals, but also transmitted virally to occupy places in books. One such example is the book recently edited by Park *et al.*[27a] In it, the titles of the second, seventh and eighth chapters conceive the phrases "… Organohalide thin films …", "Organometal halide perovskite …", and "…Organometal halide perovskites", respectively. Similarly, the book edited by Tiwari *et al.* conceives a chapter entitled "*Organometal Halide Perovskites* for Photovoltaic Applications",[27b] and so on.[27c]

An immediate question now arises: *are methylammonium lead trihalide perovskite solar cells organometallic?* Do these novel systems have any such connection with the conceptual chemistry associated with organometallics, a topic which is generally introduced to students through the undergraduate level chemistry text books? To answer these questions, let us recall some texts from undergraduate chemistry books,[28a] as well as those summarized in Wiki.[28b] As such, chemical compounds containing *at least a bond between the carbon atom of an organic molecule (or of an organic anion) and a metal*, including alkali, alkaline earth, and transition metals, and other cases,[28a] are generally treated as organometallic, and the complexes so-formed are referred to as organometallic complexes. Compounds such as transition metal hydrides and metal phosphine complexes are often treated as organometallic compounds. Other most common examples of organometallic complexes include *n*-Butyllithium, Cobaltocene, Ferrocene, Tris(triphenylphosphine)rhodium carbonyl hydride, Zeise's salt, and Trimethylaluminium, etc. Clearly, an organometallic compound traditionally yet potentially drives an inorganic metal containing system to sustain coordination with an organic moiety through a direct bonding link, [28] which, in numerous occasions, is mixed character of bonding interactions (ionic plus covalent).[29]

Thus considering these backgrounds into account, the answer to the question raised above is sharply 'No'. That is, *the $CH_3NH_3PbX_3$ (or their other mixed trihalide derivative) perovskite solar cells are neither organometallic, nor they do have any such direct connection with organometallic chemistry. We claim so as no part of the $CH_3NH_3^+$ organic cation in these specific solar cell systems is linked/bonded directly with the $Pb^{2+}$ metal core to form coordination complexes*. Evidently, if correct terminology to address the scientific content is lost, there is then no place for organometallic chemistry to be regarded as a compartmental stream of science. This means all scientific terminologies developed so far if mixed, one has then no way to separate out inorganic and coordination chemistry from noncovalent and organometallic chemistry, for example.

In fact, it is certainly not that difficult to discriminate organic-inorganic hybrid perovskites from organometallic perovskites.[30] The often cited formula $CH_3NH_3PbY_3$ that identifies the term 'methylammonium lead trihalides' is the consequence of an well-defined marriage between two monomeric subunits. These two subunits are $CH_3NH_3^+$ and $PbX_3^-$. The former is purely organic and acidic, while the latter is entirely inorganic and basic. Their unified combination facilitates significant transfer of charge between the bonding and antibonding molecular orbitals (> 0.14 *e*, larger than typical value << 0.1 *e*), forming an ion-pair of the form $CH_3NH_3^+PbX_3^-$ as likely as $Na^{\delta+}$—$Cl^{\delta-}$.[31] Because of this specific union between the two monomeric fragments scientists call the resulting system as *hybrid*, which is not predominantly driven by electrostatics only, suggested in numerous instances. [35-38] There are other potentially attractive interactions such as induction (polarization and charge transfer), and dispersion that appreciably contribute unbelievably largely to the geometrical stability as well. For instance, with the SAPT2+3/TZP level truncation such component energies for the highest light harvester $CH_3NH_3SnI_3$ are ca. –99.38, –26.61 and –11.95 kcal mol$^{-1}$, respectively, with the block geometry was energy-minimized with PBEPBE/aug-cc-

pVTZ(CH$_3$NH$_3^+$)/aug-cc-pVTZ-PP(SnI$_3^-$) using Gaussian 09, [32] where SAPT represents the Symmetry-Adapted Perturbation Theory corroborated in the PSI4 code.[33] This result clarifies the fact that the intermolecular binding interaction between the organic and inorganic species in CH$_3$NH$_3$PbX$_3$ is not just ionic as previously suggested, [35-38] but it contains significant amount of covalency.

Nevertheless, the bond metallicity can be defined by the relation $\xi_J = \rho(\mathbf{r})/\nabla^2\rho(\mathbf{r})$ at bond critical point (bcp) between bonded atomic basins in inter(or intra)molecular domains, where $\rho(\mathbf{r})$ and $\nabla^2\rho(\mathbf{r})$ are the charge density and the Laplacian of the charge density at the bcp, respectively, obtained with quantum theory of atoms in molecules. [34a] According to Jekins, [34b] $\xi_J >1$ indicates a metallic interaction present in any chemical system. However, our calculation gave a value less than unity for $\xi_J$ (i.e., $\xi_J \cong 0.6$). This result vividly signifies the fact that the I•••H–N intermolecular bonding interactions responsible to unite the two fragments CH$_3$NH$_3^+$ and PbI$_3^-$ in the CH$_3$NH$_3$PbI$_3$ complex configuration are not metallic by character. Neither the three dots in the interaction, nor the Pb–I coordinate bonds in CH$_3$NH$_3$PbI$_3$ are organometallic, as the former fragment is an organic and the latter an inorganic content. This demonstration gives the direct proof about the fact that the CH$_3$NH$_3$PbY$_3$ system should not /cannot be called/regarded as organometallic in any occasion.

Because CH$_3$NH$_3^+$ and PbY$_3^-$ are mixed in CH$_3$NH$_3$PbY$_3$ several researchers refer these and their derivatives by calling them with names such as *organic-inorganic hybrid* perovskite solar cells, or even *hybrid organic-inorganic* perovskite solar cells. We recommend to cite the CH$_3$NH$_3$PbY$_3$ systems in the TITLEs of the upcoming research articles straightforwardly as "methylammonium lead trihalide organic-inorganic hybrid perovskite solar cells", or simply "CH$_3$NH$_3$PbY$_3$ hybrid perovskite solar cells", or with analogous variants that must not confuse the reader to view these systems as how the phrase "*organometallic halide*" enforces to. Our recommended terminologies can be applicable to other trihalogen-based organic-inorganic hybrid perovskite solar cells that replace either Pb with Sn/Ge/other lead-free metal ions, or CH$_3$NH$_3^+$ with other organic cations such as in [H$_3$N(CH$_2$)$_6$NH$_3$]PbI$_4$[39a)] and [(C$_n$H$_{2n+1}$NH$_3$)$_2$PbI$_4$] ($n$ = 12, 14, 16 and 18). [39b)]

Note further that the number of research articles published to date on the trihalide-based organic-inorganic hybrid perovskite solar cells are not exactly the same as what are shown in Figs. 1a) and b). In fact, the number of actually vary with the nature of the phrase used to search the

Web of Science database. For examples, individual searches of phrases like 'organometal halide', 'organometallic halide perovskite', 'organolead halide', 'organolead iodide perovskite', '$CH_3NH_3PbI_{3-x}Br_x$ mixed halide perovskites', '$CH_3NH_3PbX_3$', '$CH_3NH_3PbI_3$', $CH_3NH_3PbBr_3$, or 'mixed halide perovskite, as well as many other variants, result in different number of peer-reviewed items published. Thus, constricting to a specific phrase such as the ones we recommended above for the title of any upcoming articles on trihalide based perovskite solar cell materials will undoubtedly assist us in identifying the exact growth of research publications in this area that will be documented in the Web of Science database, or in other database. Nonetheless, the criticism we provided above in regard to the organometallic nature of the $CH_3NH_3PbY_3$ organic-inorganic hybrid perovskite solar cells is perhaps in line with others.[30] In this latter study, the authors have also mentioned that the organic-inorganic perovskites can be characterized only with direct bonding linkage between the metal and an organic ligand. One such prototypical example in the area of perovskite solar cell is $[C(NH_2)_3M(HCOO)_3]$.[40a-b),41a] In this specific case, the organic guanidinium cation $[C(NH_2)_3]^+$ occupies the $A^+$ site, and the $HCOO^-$ formato anion replaces the halogen $X^-$ (but as an organic anion) to form the $ABX_3$ organometallic perovskites, where metal cation B = Mn, Fe, Co, Ni, Cu, and Zn. Other examples,[40c),41] as well as those involve the $[NH_4][M(HCOO)_3]$ (M = Sc to Zn) organometallic perovskites,[42] are catalogued elsewhere.

**Added note**

An initial version of the manuscript was first made available online on cond-mat arXiv (https://arxiv.org/abs/1703.09885) on March 29, 2017. The author received through private communication an acknowledgement from one of the Senior Editors of ACS Energy Letters (Professor Filippo De Angelis) mentioning that the idea perovskites are not organometallic is transmitted worldwide on April 14, 2017 through their editorial on perovskites entitled "Riding the New Wave of Perovskites" (*ACS Energy Lett.*, 2017, 2, 922–923), available via: http://pubs.acs.org/doi/full/10.1021/acsenergylett.7b00256.

**Acknowledgement**

The author sincerely thanks his research colleagues Arpita Varadwaj for reading, scientific discussions and suggestions for improving the quality of the manuscript, and Koichi Yamashita for continuous research support.

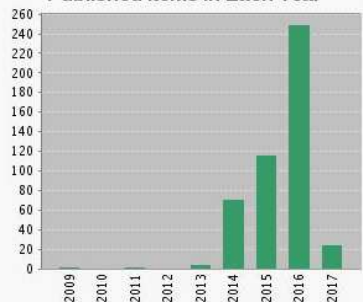
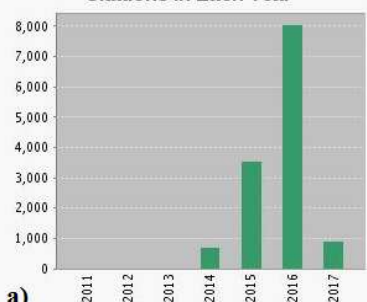
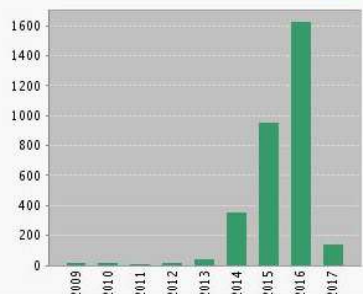
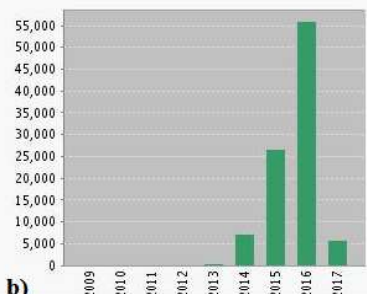
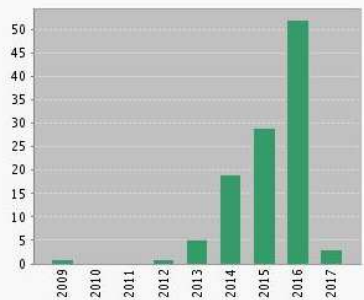
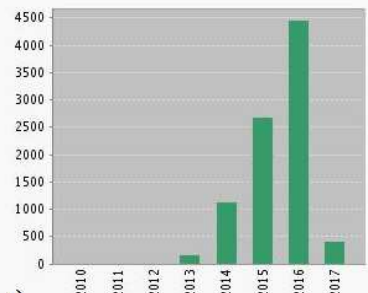
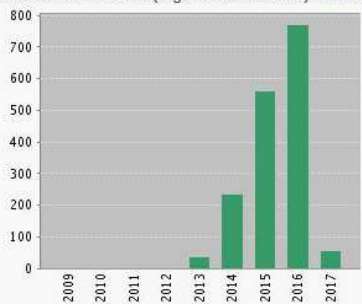
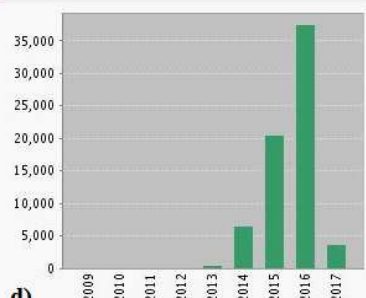

Fig. 1: Each panel (a)-(d) presents peer-reviewed research publications on trihalide-based perovskite solar cells each year (left) and their citations (right). The histograms are generated using the Web of Science Database[19] by searching keywords "halide perovskite" and "organometal halide, both as TITLE and TOPIC.